# The TaichuPix1: A Monolithic Active Pixel Sensor with fast in-pixel readout electronics for the CEPC vertex detector

T. Wu,[a,b] S. Grinstein,[b] R. Casanova,[b] Y. Zhang,[c,d,1] W. Wei,[c,d,1] X. Wei,[e] J. Dong,[f] L. Zhang,[f] X. Li,[c,d] Z. Liang,[c,d] J. Guimaraes da Costa,[c,d] W. Lu,[c,d] L. Li,[f] J. Wang,[e] R. Zheng,[e] P. Yang[a] and G. Huang[a]

[a] *PLAC, Key Laboratory of Quark and Lepton Physics (MOE), Central China Normal University,*
*152 Luoyu Road, Wuhan, 430079, Hubei, China*

[b] *Institute for High Energy Physics (IFAE), Autonomous University of Barcelona (UAB),*
*Bellaterra, 08193, Barcelona, Spain*

[c] *Institute of High Energy Physics Chinese Academy of Sciences,*
*19B Yuquanlu, 100049, Beijing, China*

[d] *State Key Laboratory of Particle Detection and Electronics, 19B Yuquan Road, Shijingshan District, Beijing, China*

[e] *School of Computer Science and Engineering, Northwestern Polytechnical University,*
*127 Youyixilu, Xi'an, 710072, Shaanxi, China*

[f] *Institute of Frontier and Interdisciplinary Science and Key Laboratory of Particle Physics and Particle Irradiation, Shandong University, Qingdao, 266237, Shandong, China*

 *E-mail:* zhangying83@ihep.ac.cn(Y. Zhang), weiw@ihep.ac.cn(W. Wei)

ABSTRACT: The proposed Circular Electron Positron Collider (CEPC) imposes new challenges for the vertex detector in terms of high resolution, low material, fast readout and low power. The Monolithic Active Pixel Sensor (MAPS) technology has been chosen as one of the most promising candidates to satisfy these requirements. A MAPS prototype, called TaichuPix1, based on a data-driven structure, together with a column drain readout architecture, benefiting from the ALPIDE and FE-I3 approaches, has been implemented to achieve fast readout. This paper presents the overall architecture of TaichuPix1, the experimental characterization of the FE-I3-like matrix, the threshold dispersion, the noise distribution of the pixels and verifies the charge collection using a radioactive source. These results prove the functionality of the digital periphery and serializer are able to transmit the collected charge to the data interface correctly. Moreover, the individual self-tests of the serializer verify it can work up to about 3 Gbps. And it also indicates that the analog front-end features a fast-rising signal with a short time walk and that the FE-I3-like in-pixel digital logic is properly operating at the 40 MHz system clock.

---

[1]Corresponding author.

# Contents

---

## 1. Introduction

The Circular Electron Positron Collider (CEPC) has been proposed for high-precision measurements of the Higgs boson. The tracking detector is critical to achieve an excellent impact parameter resolution. Besides, it must be able to withstand ionizing radiation to ensure stable operation in the high luminosity collision environment of the CEPC. Its performance directly determines the physical output of the CEPC experiment. Monolithic Active Pixel Sensors (MAPS) are being investigated for the CEPC vertex detector due to their high granularity, high speed, low material budget, low power consumption and radiation tolerance.

The R&D program to study the viability of MAPS for the CEPC vertex detector started in 2016 in the framework of the MOST1 project (supported by the Ministry Of Science and Technology of China). Several prototypes were produced in order to study the performance of different architectures. JadePix-1 [1] and JadePix-2 [2] were designed with different pixel sizes to investigate the large area and high resolution. At the same time, another prototype, the MIC4 [3], was designed to study a data-driven readout architecture. The results of the MOST1 are promising but distant from a complete prototype that is needed for the vertex detector, so the MOST2 project was launched in 2018, in order to develop a fully functional pixel sensor and implement the baseline structure of the inner tracking detector. The requirements for the full-scale pixel sensor prototype are summarized in table 1. The dead time is the maximum duration to read out one double column. This time should be less than 500 ns to achieve a detection efficiency higher than 99%.

The device cross-section is shown in figure 1(a). The sensor and readout electronics are integrated on the same silicon bulk. The p-n junction is reverse biased to create a depletion region around the n-well. When a particle crosses the bulk, pairs of electron-holes are created by ionization. The electrons diffuse towards the n-well collecting electrode. Driven by the external electric field, electron-hole pairs created inside the depletion region or arriving by diffusion to it are collected by drift. The readout electronics are integrated inside a deep p-well that surrounds the collecting electrode with an octagonal shape, figure 1(b). The diameter of the n-well electrode is 2 μm and the n-well to p-well spacing is between 2 and 3 μm, while the total dimension of the pixel in one direction is 25 μm (see figure 1(d)). In terms of the area of one pixel, the size of the collection diode with spacing around is 8.6×8.6 μm², the layout area for the analog front-end is 8.2×25 μm², and the rest of the space is reserved for the in-pixel digital logics and bus routing. The thickness of the epitaxial layer is 25 μm. The sensor reset is provided by a diode which makes the sensor compact. The schematic of this method is illustrated in figure 1(c).

This paper is focused on the first MAPS prototype named TaichuPix1 developed within the MOST2 project. It has been implemented in a 0.18 μm CMOS Image Sensor (CIS) process. It includes a 192×64 matrix with a pixel size of 25×25 μm² [4]. Regarding the previous pixel sensor prototypes for the CEPC vertex detector, the TaichuPix1 is the first to satisfy both pixel size and readout speed. The analog front-end has a fast-rising edge with a short time walk, since the bunch spacing of the CEPC colliding beams are 680 ns, 25 ns, and 210 ns, respectively, for the Higgs, Z and W operations [5]. A 40 MHz reference provides the operating clock to the peripheral readout logic. The in-pixel readout logic that works at the main clock of 40 MHz has a very limited area. At the same time, the digital periphery and a serializer with an internal PLL (Phase Lock Loop) are designed to be compatible with high data rate readout architecture [6].

The overall architecture of TaichuPix1 is described in section 2. A description of the test setup is given in section 3 and experimental results are shown in sections 4 and 5. Some limitations of this prototype are addressed in section 6. Finally, a summary and the conclusions are presented in section 7.

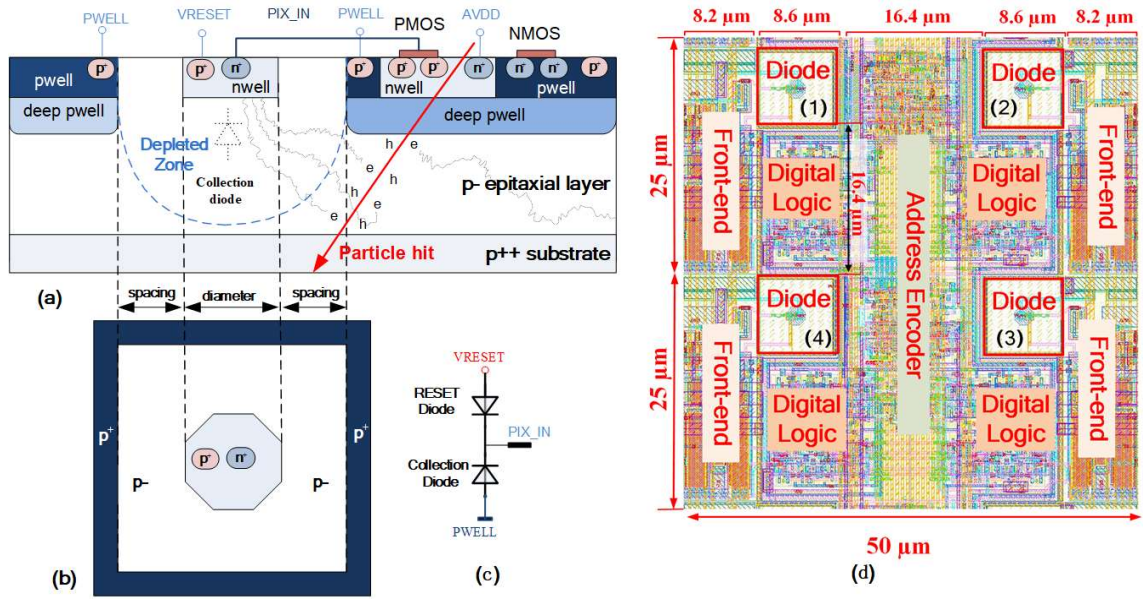

**Figure 1.** Cross section(a), top view(b), RESET schematic(c) and four adjacent pixels (d) of TaichuPix1

Table 1. Main design specifications of the pixel sensor in this work

| Parameter | | Value |
|---|---|---|
| Single point resolution (μm) | | 3~5 |
| Bunch spacing (ns) | | Higgs:680; W:210; Z:25 |
| Hit density(hits/event/cm$^2$) | | Higgs:2.4; W:2.3; Z:0.25 |
| TID radiation hardness (Mrad) | | > 1 |
| NIEL radiation hardness (1 MeV $n_{eq}$/cm$^2$) | | $2\times10^{12}$ |
| Pixel size (μm$^2$) | | 25×25 |
| Operating frequency (MHz) | | 40 |
| Data rate | Triggerless (Gbps) | 3.84 |
| | Trigger (Mbps) | 160 |
| Dead time (ns) | | < 500 |
| Power density (mW/cm$^2$) | | < 200 |

## 2. The overall architecture of TaichuPix1

The TaichuPix1 ASIC (shown in figure 2) is divided into two main blocks, the pixel array and the periphery circuitry. The former contains 192 columns (96 double columns) with 64 rows, two columns of pixels share the same data bus for readout (see address encoder region in figure 1(d)). The latter includes two-level FIFOs, a readout controlling monitor, two types of DACs (Digital to Analog Converter), and a serializer.

### 2.1 Pixel Design

As shown on the right upper corner of figure 2, each pixel consists of a sensing diode and an analog front-end surrounded by the in-pixel digital logic. The analog front-end includes a preamplifier followed by a shaper and a hit discriminator based on the ALPIDE [7] architecture. In this work, each pixel has the same analog front-end, which has been optimized for a faster-rising edge. Two new digital fast in-pixel readout logics have been designed: an FE-I3 [8] like scheme (scheme1 in figure 2) and ALPIDE-like scheme (scheme2 in figure 2) benefited from the AERD (Address Encoder and Reset Decoder) [9] structure. In the first scheme, the in-pixel digital electronics follows the FE-I3 design. However, due to area limitations, the address ROM has been replaced by a pull-up and pull-down network matrix, and the timestamp is not stored in-pixel but at the end of the column. In the ALPIDE like scheme, the pixel cell structure is the same. Still, a dynamic edge-triggered flip-flop has replaced the hit storage registers in order to prevent reading out one hit repeatedly before the analog front-end resets. In addition, the priority block AERD has been modified to boost its speed to 40 MHz with respect to [9]. There are two different methods to implement the AERD readout in [9]. One is the standard architecture to read out with the rising edge of the clock, and the other is the boosting speed one, the extra address enable signals are added to read out with both rising and falling edge. The latter one has been designed, but the experimental results are not as good as expected.

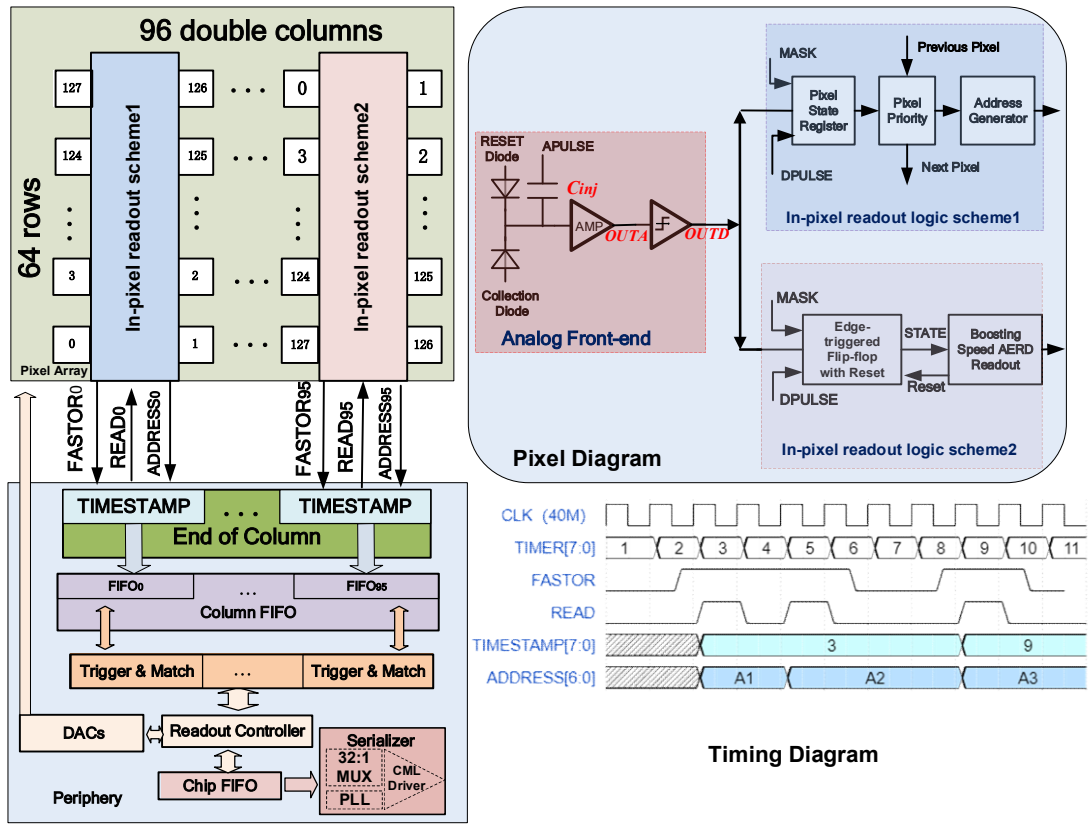

**Figure 2.** The chip architecture of TaichuPix1, Pixel diagram and Timing diagram are on the right.

The pixel matrix is subdivided into two matrices of 96×64 pixels, each with a different in-pixel readout digital logic (FE-I3 and ALPIDE like). However, both schemes employ the same double-column drain architecture [10] in the peripheral part of the chip, as shown in the bottom part of figure 2. the same peripheral readout logic was implemented at the end of the column. More details about the readout architecture can be found in [4].

The region for in-pixel digital readout logic is shared by two columns so that the cross-talk between analog biasing lines and digital buses is reduced. At the same time, it saves space for the routing of the address encoder (figure 1(d)). The priority logic arbitrates the pixel readout, with the topmost pixel having the highest readout priority. Both schemes have the same coding sequences in a serpentine approach for each double column, the coding order for the FE-I3-like scheme starts from 127 to 0 and the queue for the ALPIDE-like scheme is from 0 to 127.

## 2.2 Periphery circuitry

At the EoC (End of Column), the circuitry has a counter running at 40 MHz to generate the timestamp with a step of 25 ns (see timing diagram in figure 2). The timestamp is recorded at the rising edge of FASTOR, and it will match the hit address at the EoC. If two hits arrive in a short time in the same pixel, they will be merged as one. If the hits are in different pixels, they will be read out with the same timestamp.

TaichuPix1 can operate in a trigger mode where the data that matches the timestamp are stored temporally for each double column in the first level FIFO (fifo_column). Then the data matched with a trigger is passed to the second level FIFO (fifo_chip), and finally serialized.

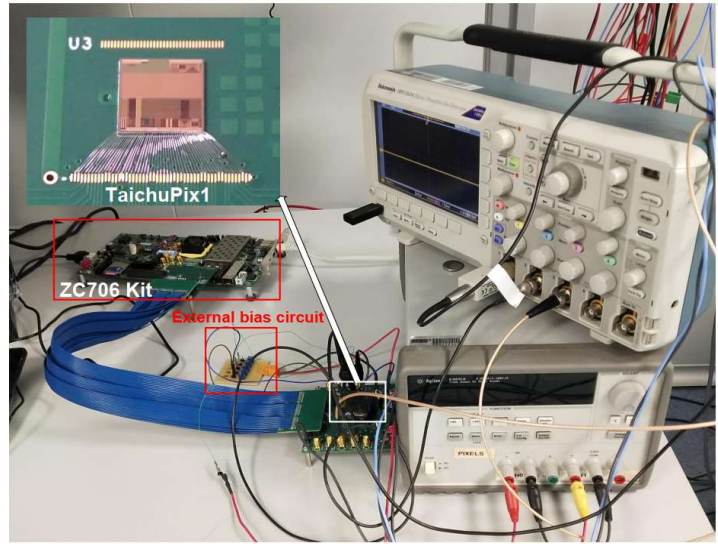

**Figure 3**. Test Platform for TaichuPix1

fifo_column can store 10 μs data which is big enough to satisfy the trigger latency at most of 6 μs. For example, if the trigger latency is set to 3 μs and the counter starts from 10, only when the data timestamp is equal to 130 will it be readout. There is a maximum window of 175 ns to record matched timestamps, while uncertain timestamps will be arbitrated in the trigger discriminating logic. The average trigger rate is 50 kHz, and the trigger latency is supposed 3~6 μs. The required data rates correspond to the physics case scenarios described in [6]. An interface of 160 Mbps in trigger mode is capable of meeting all these requirements. TaichuPix1 can also operate in triggerless mode. In this case, all the readout data are preserved before buffering to the output.

In order to implement a high-speed serial transmission, the clocks are managed by a PLL integrated into the serializer, which can offer several Giga-bps data rate capabilities. In addition, the analog biasing is supported by the 8-bit current DAC and 10-bit voltage DAC, and the external supply can optionally control the reference bias for the DACs.

## 3. Test platform for TaichuPix1 Chip

The test platform for the TaichuPix1 (shown in figure 3) includes a dedicated board where the device is wire bonded, a Xilinx ZC706 FPGA evaluation board, an adapter board with an FMC connector, an oscilloscope, a DC power supply and a PC. The ZC706 evaluation board is the main module to configure the chip and transmit the readout data from the chip to the PC. As it is shown in figure 4, the ZC706 SoC consists of an integrated processing system (PS) and programmable logic (PL) on a single die. The PS is composed of an ARM processor with different peripherals connected through an AXI bus. One of these peripherals is an ethernet port which is used to communicate the ARM processor with a PC. On the side of PL launches an SPI interface to communicate with the TaichuPix1, which deserializes all the data stored in fifo_chip first before sending it to the PC via Ethernet port. In this work, the main clock of the chip is 40 MHz, the speed of SPI is 10 MHz, and the frequency of the data interface is set to 160 MHz. Besides, in order to provide enough power to the chip, TaichuPix1 is power-up directly from an external DC power supply. Considering the cable transmission resistance loss, this can be offset by increasing the input voltage.

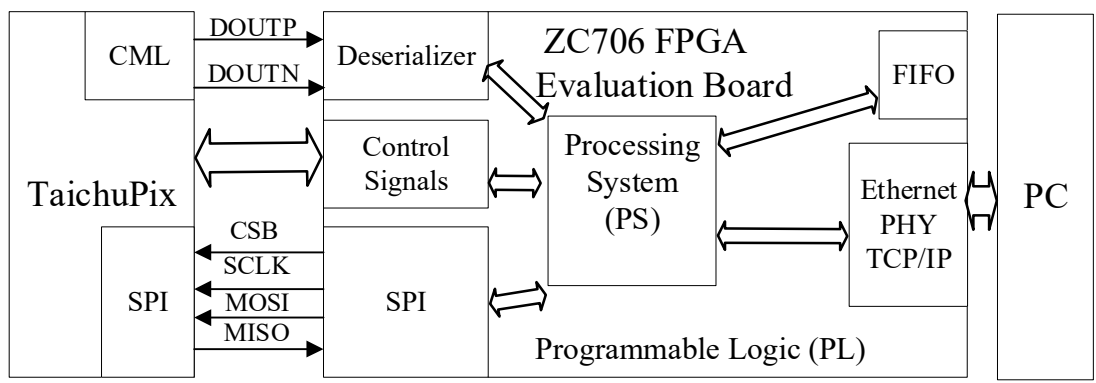

**Figure 4.** Setup system diagram based on ZC706 for TaichuPix1

## 4. Preliminary results of analog front-end and in-pixel readout electronics

The analog front-end is calibrated by injecting charge with an impulse response switch circuit. It is switched on by the APULSE input. The $C_{inj}$ (see figure 2) is an internal calibration capacitor used to verify the analog front-end. The amount of charge is adjusted via the external step voltage input ($\Delta V$). By recording the number of times that the injected charge is above the threshold and generates a signal, one can obtain the response ("s-curve") of each pixel. The temporal noise (TN) can be extracted by fitting the "s-curve." The TN distribution and fixed pattern noise (FPN) can be measured by analyzing the "s-curves" of all the pixels.

The analog front-end electronics is tuned through several DACs. The CMOS process determines the value of the calibration capacitor of each pixel. For the preliminary test presented in this section, PWELL is connected to the 0 V (see figure 1), however, this value should be -6 V in a normal condition. The charge to voltage factor from the simulation is around 0.9 mV/e$^-$, but it has not been calibrated. In order to make the testing results more accurate, all of the charges will be presented in the form of injected voltage. Figure 5 shows the measured response of the preamplifier output (OUTA in figure 2) to different injected voltages. The OUTA waveform of 400 mV injection looks different since a clipping mechanism is implemented to limit the pulse duration for large input signals [7]. The peaking time measured from the $C_{inj}$ is less than 400 ns, while this value is faster than the prototype of ALPIDE (2 μs) and MIC4 (1 μs). The delay of the leading edge roughly measured by an oscilloscope brought a time walk of 57 ns (see figure 12). The transient waveforms show that the analog front-end is able to respond to the different charges and more charges lead to a faster-rising edge. The s-curve of one pixel (column 61, row 6) is fitted on the right side of figure 5, the mean value of Gaussian fit corresponds to a threshold of 225.2 mV, and the sigma of the fit is the TN of 10.3 mV.

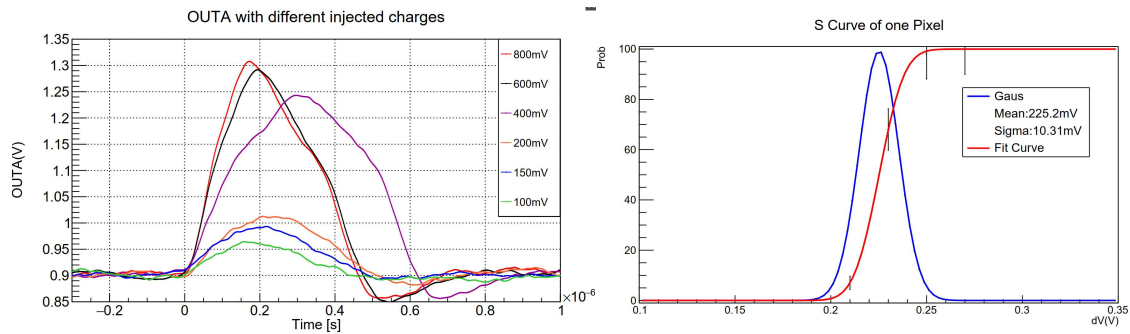

**Figure 5.** Pixel front-end transient plots under different injected charges and S-curve fit of one pixel

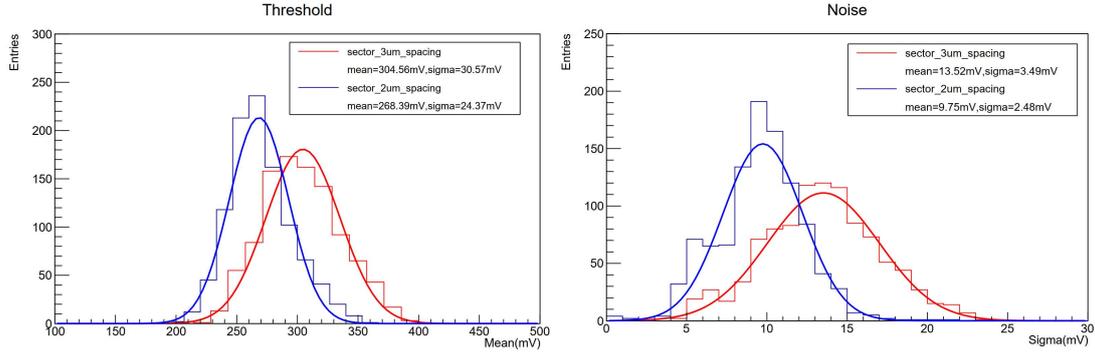

**Figure 6.** Threshold distribution and noise distribution of 16 columns.

Figure 6 shows the FPN of the FE-I3-like matrix. There are two different sensor geometries designed for the FE-I3-like pixels, the left half of the matrix is sector_3μm_spacing (columns 0 to 47) and the right half is sector_2μm_spacing (columns 48 to 95). As shown in figure 1(b), the spacing of sector_3μm_spacing is 3 μm and for sector_2μm_spacing is 2 μm. The threshold dispersion is shown on the left side of figure 6 and the right side is the noise distribution. In the noise measurement, the charge was injected by the step voltage through the $C_{inj}$ in each pixel. Thus, the noise value obtained in such a way indeed is the sum of both the fluctuations on the injected signal and the fluctuations on the threshold. The pixels of sector_2μm_spacing behave better in terms of threshold and noise. The average threshold for sector_3μm_spacing is around 304.6 mV while the value for sector_2μm_spacing is about 268.4 mV, and the FPN is 30.6 mV and 24.4 mV, respectively, while the average TN is 13.5 mV and 9.7 mV, respectively. At the given threshold of figure 6, the noise rate is around 15.9 hit/cm$^2$/min. Since the global threshold of the pixel array is adjusted by the same DACs, there is no way to optimize the dispersion by tuning each pixel.

## 5. Fully functional characterization with a radioactive source

The full functionality of the TaichuPix1 was verified using electrons from a $^{90}$Sr radioactive source. The data inside the fifo_chip can be read out directly via the SPI interface while the chip operates at the debug mode in the digital periphery circuitry. In this mode, the serializer is switched off. Figure 7 indicates the overall test process. The $^{90}$Sr (3.941 kBQ, from 15-Aug-2011) radiation source is placed directly on the top of the chip (separated by 1cm approximately). The analog front-end is configured to a proper value by the on-chip DACs. In particular, the settings are the same as shown in section 4. The measured response of the preamplifier to electrons is shown in figure 8. It proves that the analog front-end is able to amplify the charge generated by the beta source $^{90}$Sr. Comparing to figure 5, the signals in figure 8 show different shapes because the input signals for figure 5 are injected voltages to the front-end through the $C_{inj}$.

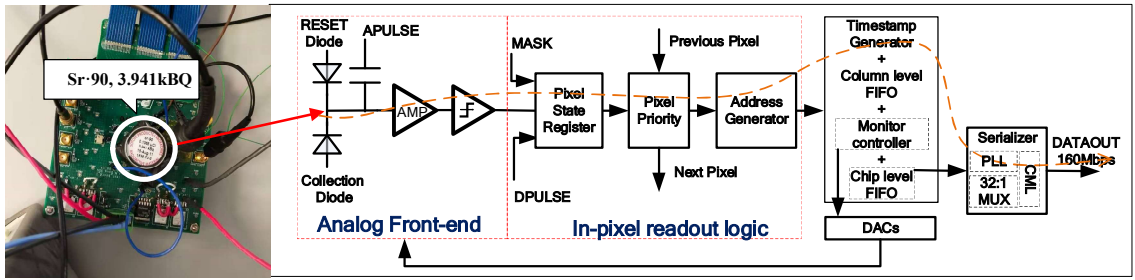

**Figure 7.** Signal flow of the TaichuPix1 under $^{90}$Sr exposure

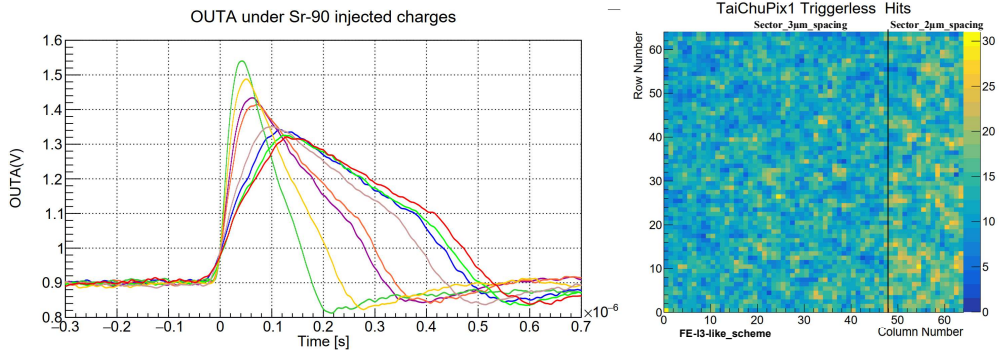

**Figure 8.** Analog front-end output and hit map under $^{90}$Sr exposure at triggerless mode.

The feasibility of a high-speed serializer is also taken into consideration. The serializer is composed of a ring-oscillating PLL, a 32:1 multiplexer, and a CML driver. The system reference clock is 40 MHz. The measured frequency locking range of PLL is from 320 MHz to 2.68 GHz. The highest average data rate of the detector is 3.84 Gbps, which agrees with the stringent average hit rate of the W bosons in triggerless mode [6]. The preliminary self-tests by using an internal PRBS-$2^7$ (Pseudo-Random Binary Sequence) generator as the data source of the serializer show it is hard to work at about 4 Gbps limited by the technology and driving capability, but can work up to 3 Gbps (see figure 9). Each pixel is recorded with a 32-bit word in the following format (see figure 10). It has 1-bit data available flag ("1" means data valid, "0" means data invalid), 8 bits timestamp, 19 bits pixel address information, and 4 bits data compression pattern [6]. A reference pattern of data has to be added to the data output to be able to decode it. Otherwise, it is difficult to synchronize the data as it only relies on the 1-bit data available flag.

Figure 8 shows a hit map of the device exposed to the radioactive source for 12 hours, where the data is taken from the serializer. The measurement corresponds to 64 columns of the FE-I3-like pixels of the chip, operated at 160 Mbps. The device is working in the triggerless mode, and the measurement setup is based on figure 7. The areas of the chip with slightly different thresholds (indicated in figure 8 as sectors with 3μm and 2μm spacing) result in slightly different occupancies. The average threshold for the sector_3μm_spacing (2μm) is 317 mV(280 mV). For the sector_3μm_spacing, the chip runs for 3 hours without any injection; three hits are recorded, converted to the noise rate of 0.65 hit/cm$^2$/min.

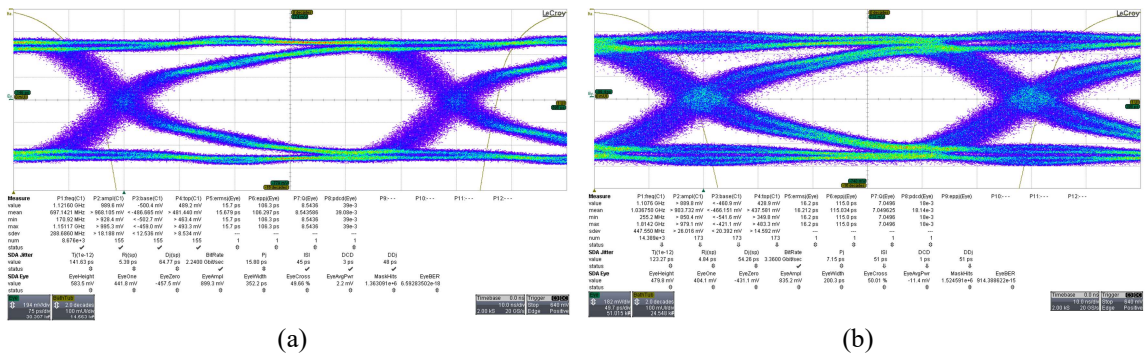

(a)                                    (b)

**Figure 9.** Eye diagrams of PLL at self-test mode with the speed of 2.24 GHz(a) and 3.36 GHz(b)

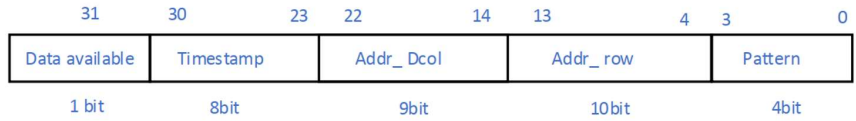

**Figure 10.** Output data format

The TaichuPix1 matrices are divided into 3 fast readout groups with 64 columns in each group. According to the peripheral readout approach designed in the chip [6], only one data group can be read out when hits arrive. That is why figure 8 presents the hit map of one of the groups so that each column has the same priority to be read out. However, in this operation scheme, data is lost, as it becomes apparent when comparing to the triggered mode, shown below. On the one hand, when too many hits are coming simultaneously, only part of the events can be read out, and the hits arriving later will be lost. On the other hand, the triggerless mode is used to record any hit above the threshold, thus it will record many invalid data (due to hot pixels), which saturates the readout resources, resulting in the $^{90}$Sr events being lost. This leads to a relatively low event count (lower than 40 per pixel) even for 12 hours of exposure to the $^{90}$Sr source.

Besides the triggerless mode, trigger mode is also verified with the external scintillator trigger setup (see figure 11). The TaichuPix1 is placed between the scintillator and the $^{90}$Sr source, as illustrated in figure 11(a). A high voltage supply powers up the scintillator with -900 V. The signal of the scintillator is connected to a discriminator board with a NIM (Nuclear Instrument Modules) level output signal. A level adapter board is used to covert the NIM to TTL (Transistor-Transistor Logic), which is used for triggering the readout system. The $^{90}$Sr beta electrons travel through the chip and reach the scintillator. The chip is operated in trigger mode, with a latency of 3 μs. The uncertain data window is set to 175 ns, which can read out trigger data with the error range of 7 timestamps.

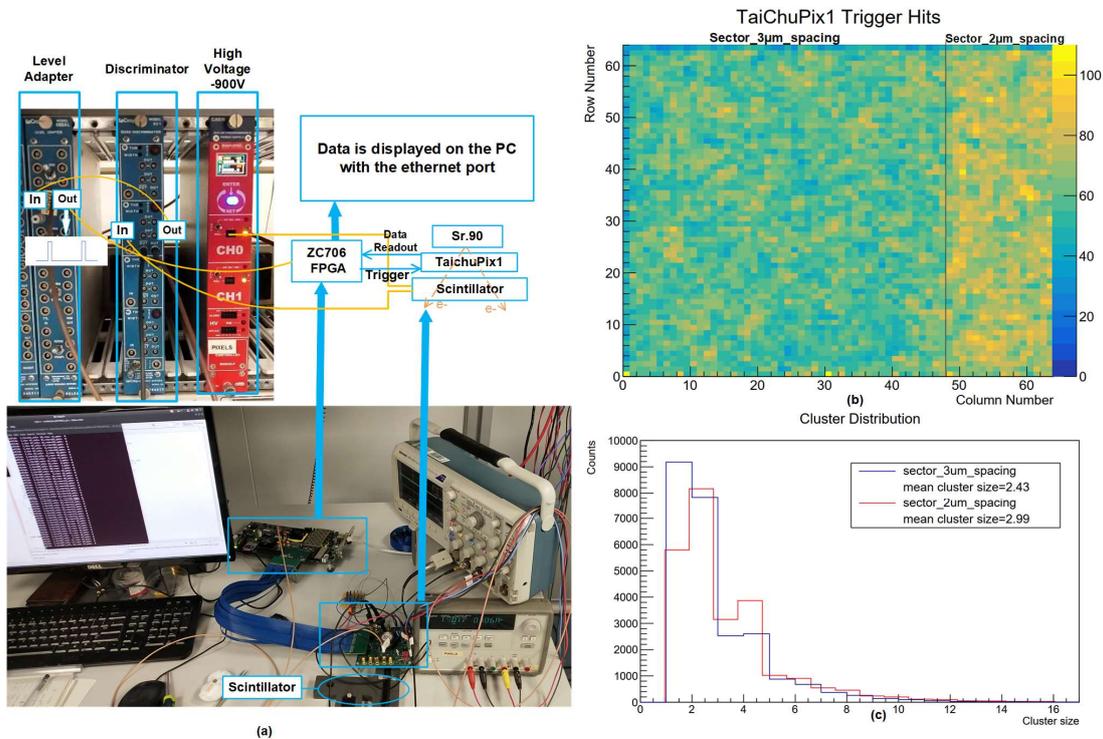

**Figure 11.** Trigger setup (a), hit map (b) and cluster size(c) under $^{90}$Sr exposure in trigger mode

Considering the fast group readout priority, only one of the groups is turned on. Using these trigger settings, the hit map with a scale of about 100 hits per pixel for 4 hours run is drawn in figure 11(b). The shape gradient is matched with the threshold dispersion between sector sector_3μm_spacing and sector_2μm_spacing. And the pixels at the edge of the matrix have a lower hit rate due to the dummy pixels, which are only placed at the bottom of the chip. Neither the left nor the top sides have dummy pixels. In addition, the cluster size can be analyzed with the information of the timestamps and the hit addresses. The cluster distributions under the exposure to $^{90}$Sr of these two sectors are shown in figure 11(c). Each sector equally calculates 16 columns. From the histogram result, it can be concluded that the lower threshold leads to a larger cluster size. The average cluster size of this hit map is around 2.6 hits.

The hit map shows qualitatively that the sensor can successfully detect traversing charged particles. Both triggerless and trigger mode are verified. It proves the functionality of the peripheral readout logic design.

## 6. Limitations of TaichuPix1

As mentioned in section 4 the results included in this paper were obtained with the PWELL connected to 0 V, that is, with no reverse bias in bulk. Applying a -6 V bias on the chip substrate is designed to decrease the sensor capacitance, leading to a lower threshold. However, the chip measured with a -2 V substrate bias does not show an obvious decrease on the threshold. It was

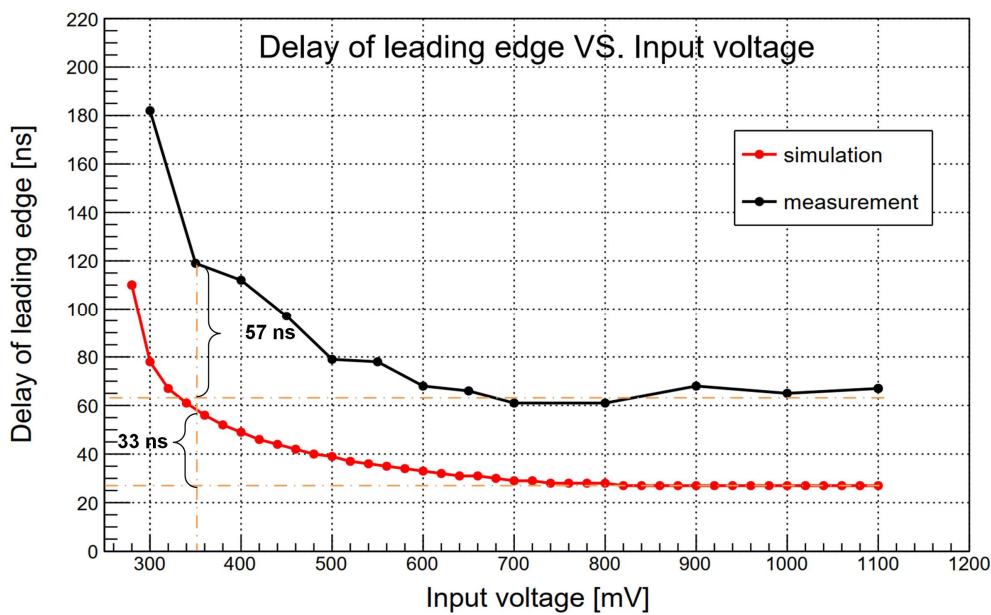

**Figure 12.** Delay of leading edge vs OUTA input voltage. The evaluated input capacitance is set to 5 fF while the PWELL is connected to the 0 V. The time walk in this threshold from 350 mV input voltage to 1100 mV is around 57 ns, while this value to the simulation is 33 ns. The test method is to set the rising edge of APULSE as the time reference, recording the delay between APULSE rising edge and OUTA leading edge.

Table 2. Power consumption evaluation

| Parameters | | TaichuPix |
|---|---|---|
| Power consumption | Analog front-end | 130 mW/cm$^2$ |
| | In-pixel digital logic | FE-I3-like scheme[a]: 6.15 mW/cm$^2$ |
| | | ALPIDE-like scheme[a] : 15.73 mW/cm$^2$ |
| | Periphery Circuitry | Trigger mode[a][6] : 25~30 mW/cm$^2$ |
| | | Triggerless mode[a][6] : 80~140 mW/cm$^2$ |
| | Serializer | Trigger: 153 mW  Triggerless: 234 mW |

a. Estimation condition: The chip sensing active area is 3.2768 cm$^2$, the pixel array is 512×1024, the pixel pitch is 25 μm, the bunching spacing is 25 ns, the cluster size is 3 pixels, the epitaxial layer is 25μm, the average trigger rate is 50 kHz and the average hit time is 8.3 μs per column.

observed that applying a reverse bias will sometimes damage the chip through a process that is not fully understood. Therefore, zero bias voltage is used to measure the threshold and noise for the moment. On the other hand, the threshold can be adjusted by varying the front-end bias parameters.

The time walk performance of the analog preamplifier is shown in figure 12. As mentioned in section 4, different injected voltages to $C_{inj}$ will lead to a different leading edge of OUTA. The rising edge of APULSE is set as a timing reference, then recording the delay of the leading edge of OUTA, a curve versus input voltage can be drawn out. The time walk at the 268.4 mV threshold (roughly 300 e$^-$) of the preamplifier measured by an oscilloscope is around 57 ns, corresponding to the input voltage from 350 mV (~390 e$^-$) to 1100 mV(~1220 e$^-$), while the simulated value is 33 ns. Although the analog front-end is optimized for a fast-rising edge and a short time walk, the standard CIS process adopted may not satisfy the speed requirement imposed by the CEPC completely. As the pixel structure in figure 1 shows, the charge collection time would be larger if the charge is generated in the non-depleted part of the pixel. A modification process [11] with a full depletion epitaxial layer will be considered in a future prototype.

The power consumption is another limitation of this prototype since the chip operates with the same principle of ALPIDE architecture, but with a faster timing performance, at the expense of increasing the power consumption (~200 mW/cm$^2$ while for the ALPIDE chip is around 35 mW/cm$^2$ [12]). More details of the full matrix power consumption evaluation can be found in table 2. The trade-off between fast timing and low power consumption will be addressed in the follow-up research.

As for the in-pixel readout logic, two different schemes were designed in parallel. However, this work only presented the experimental results of the FE-I3-like part due to the fact that the ALPIDE-like scheme has some problems with the matrix readout and due to the high-impedance state that exists in data bus at these columns. Sometimes this state will introduce an uncertain state to the digital logic and generate random data, resulting in only the top priority columns being read out correctly. One possible solution is to use to latch to pull up/down the uncertain state. Moreover, the readout timing for boosting speed AERD is not completely the same as FE-I3-like scheme, but they have employed the same column level readout architecture. The solution could be to design a new digital periphery readout dedicated to the ALPIDE-like scheme.

## 7. Summary

A functional monolithic pixel sensor prototype has been designed for the CEPC vertex detector. The TaichuPix1 is a small-scale prototype to verify the feasibility of implementing a pixel size of 25×25 μm$^2$, together with a column drain readout architecture. It presents an in-pixel analog front-end circuit that has been optimized to have a fast-rising edge and short time walk. Two parallel in-pixel readout schemes based on ALPIDE and FE-I3 were designed. Improvements were made in order to meet the requirements in terms of area and speed. The experimental characterization has been presented, including the response to charge via the APULSE injection mechanism, which resulted in an achievable threshold value of 268.4 mV (roughly 300 e$^-$), its dispersion with a typical dispersion FPN of 24.4 mV and a noise distribution with a typical TN of 9.7 mV for the pixels based on FE-I3-like scheme. The results of the laboratory measurement using a radioactive source ($^{90}$Sr) were shown. Both trigger and triggerless mode results are presented. These results prove the functionary of the digital periphery and the serializer. And it also indicates that the analog front-end features a faster-rising edge with a short time walk. The area-limited FE-I3-like in-pixel digital logic is properly operating at the 40 MHz system clock.

The TaichuPix1 is a functional demonstrator chip with small pixel size and a fast readout speed. The functional architecture based on the FE-I3-like scheme was verified, and the results are valuable for implementing a full-scale prototype in the future.

## Acknowledgments

The research was supported and financed in part by the National Key Research and Development Program of China under Grant No.2018YFA0404302, No.2016YFA0400404 and No.2016YFE0100900, National Natural Science Foundation of China under Grant No.11420101004, No.11605071, No.11835007, No.12075142 and No.11835008. Shandong Provincial Natural Science Foundation No. ZR2020MA102. And it was also partially funded by the Spanish Government, under grants FPA2015-69260-C3-2-R and RTI2018-094906-B-C21, and by the H2020 project AIDA-2020, GA no. 654168. Tianya Wu was sponsored by the China Scholarship Council CSC201806770048 and the Huabo development plan of Central China Normal University. Thanks for the strong supports from the IFAE engineering department, especially from Mr. C. Puigdengoles and Mr. J. García Rodríguez. Thanks to colleagues Stefano Terzo, Chiara Grieco, Weiping Ren, and Kai Jin for their valuable and constructive suggestions.